\documentclass[12pt]{iopart}
\usepackage[dvips]{graphicx}
\usepackage[dvips]{color}
\usepackage{textcomp}
\begin{document}

\title[The pressure dependence of the magnetic ordering transition in Na$_x$CoO$_2$]{The magnetic field and pressure dependence of the magnetic ordering transition in Na$_x$CoO$_2$ (0.6$\leq$x$\leq$0.72)}
\author{J Wooldridge, D M$^c$K Paul, G Balakrishnan and M R Lees}
\address{Department of Physics, University of Warwick, Coventry, CV4 7AL, United
Kingdom.}
\ead{J.Wooldridge@warwick.ac.uk}

\begin{abstract}
We have measured the magnetic field (H$\leq$90 kOe) and pressure (P$\leq$10 kbar) dependence of the magnetic ordering temperature, T$_{mag}$, in single crystal samples of Na$_x$CoO$_2$ for a range of Na concentrations (0.60$\leq$x$\leq$0.72). We show that in zero field, T$_{mag}$ remains constant with decreasing x before magnetic order disappears at x=0.65. Heat capacity and magnetization data show that for x=0.70, T$_{mag}$ is unchanged in an applied field. In contrast, magnetization data collected under hydrostatic pressure show that T$_{mag}$ increases from 22.0 K at 1 bar to 25.4 K at 10 kbar. This rise is at odds with the behaviour expected for model spin density wave systems.
\end{abstract}
\pacs{74.62.Fj, 75.30.Fv, 75.30.kz, 75.40.Cx, 75.50.Ee}
\submitto{\JPCM}
\maketitle

\section{Introduction}
Cobalt oxide systems have attracted considerable attention because of their interesting magnetic and transport properties. One system of particular note is Na$_x$CoO$_2$. This material displays a diverse range of characteristics including spin-entropy enhanced thermopower~\cite{wang}, charge and spin ordering~\cite{foo} as well as strong electron correlations~\cite{ando}.
\par
The structure of Na$_x$CoO$_2$ depends strongly on x. For 0.5$\leq$x$\leq$0.85 the system is hexagonal (P6$_3$/mmc) with layers of CoO$_6$ edge-sharing octahedra stacked along the \textit{c}-axis separated by layers of Na ions that are located on two partially occupied sites. The details of the structure depend on how these sodium ions are arranged in the charge reservoir layer and how the CoO$_2$ layers respond structurally to changes in the electron count and the Na ion distribution. Using the terminology adopted by Huang et al.~\cite{huang} we note that for 0.5$\leq$x$\leq$0.74, Na$_x$CoO$_2$ has an H1 structure. As x increases the \textit{c}-axis lattice parameter decreases as a result of a decrease in the thickness of the NaO$_2$ charge reservoir layer that is not offset by an increase in the CoO$_2$ layer thickness (and the Co-O bond lengths). \textit{a} increases slightly over the same x range. For 0.76$\leq$x$\leq$0.82 there is an H2 phase that is separated from H1 by a narrow two phase region around x=0.75. In the H2 phase both \textit{c} and \textit{a} continue to vary with x as for H1, although the decrease (increase) with x in the NaO$_2$ (CoO$_2$) layers is more pronounced.
\par
Sugiyama et al.~\cite{sugiyama} have reported that T$_{mag}$, has a dome shaped dependence with x, although only two compositions with magnetic order at temperatures of 19 K (x=0.9) and 22 K (x=0.75) were shown. On the other hand, Sakurai et al.~\cite{sakurai} who have studied polycrystalline samples with 0.7$\leq$x$\leq$0.78, suggest that T$_{mag}$ is constant across this region of the phase diagram, that T$_{mag}$ remains unchanged in fields of up to 90 kOe, and that there is no difference in the transition temperature for data collected while heating or cooling the sample. This is compatible with a second order phase transition into a magnetically ordered state with an AFM nature~\cite{me}. A study of x=0.85 single crystals~\cite{luo} has shown that T$_{mag}$ exhibits a field dependence. In low fields, T$_{mag}$ is reported to decrease from 18.5 K in zero applied field to 17.5 K in an applied magnetic field of 100 kOe, as the AFM correlations are suppressed. T$_{mag}$ then increases with H for higher fields. A metamagnetic transition to a FM state is seen in the low temperature (T$\leq$15K) M versus H data at 100 kOe leading to the suggestion that above this applied field the FM correlations are strengthened, thus increasing T$_{mag}$. Neutron scattering data for Na$_x$CoO$_2$ (x=0.75 and 0.82) are consistent with an A type-antiferromagnetic ordering~\cite{bayrakci,helme,boothroyd}. The exchange in Na$_x$CoO$_2$ appears to be surprisingly isotropic with FM intra plane and AFM inter plane exchange constants that are similar in magnitude. A combined EPR and NMR study of samples with 0.65$\leq$x$\leq$0.75 has indicated that at temperatures below T* (220-270 K) there is a phase separation of the Na$_x$CoO$_2$ material into metallic non magnetic regions and insulating magnetic regions containing localized S=$\frac{1}{2}$ Co$^{4+}$ magnetic moments~\cite{carretta}. This picture is consistent with the small magnetic entropy associated with the anomaly seen in the heat capacity data around T$_{mag}$~\cite{me}. There is a Coulomb penalty to pay in such a situation and the mechanism driving the phase segregation is not clear. There have been conflicting reports regarding phase separation from muon spin spectroscopy studies, with some authors reporting magnetically ordered volume fractions of between 20$\%$ and 50$\%$ while others have reported bulk magnetic order for 100$\%$ of the sample volume~\cite{carretta,bayrakci2,mendels,sigiyama}. The spin wave dispersion curves derived from neutron scattering data appear incompatible with a random distribution of localized Co$^{3+}$ and Co$^{4+}$ ions carrying S=0 and S=$\frac{1}{2}$ spins respectively. Within the phase segregation picture, the sharp spin modes require that the in-plane clusters of FM Co$^{4+}$ ions be aligned vertically above one another over many layers. A coherent picture of the nature of the magnetic order in Na$_x$CoO$_2$ is yet to emerge. A scenario in which the charge disproportionation on the Co atoms is small, with the magnetic ordering corresponding to a spin-density wave for x greater than $\sim$0.7 with A type antiferromagnetic order at x$\sim$0.8 seems to provide the most convincing explanation for the data collected to date.
\par
Here we present measurements on single crystal samples of Na$_x$CoO$_2$ (0.60$\leq$x$\leq$0.72) examining how the application of both magnetic field and pressure influence the magnetic order in Na$_x$CoO$_2$. To date there have been no reports on how the magnetic properties vary with the application of pressure. Such studies are problematic because the signature of magnetic order in Na$_x$CoO$_2$ is weak in the data collected using many of the techniques most commonly employed in combination with pressure cells such as ac susceptibility, resistivity, and neutron diffraction measurements. Here we are able to carry out such a study by measuring the dc magnetization under pressure using a pressure cell designed for use with a SQUID magnetometer.
\section{Experimental details}
Polycrystalline starting materials of Na$_{0.75}$CoO$_2$ were prepared by mixing Na$_2$CO$_3$ and CoO powders with a Na:Co ratio of 0.75:1. The well-mixed powders were calcined at 750 $^{\circ}$C for 12 hours and then reground and reacted at 850 $^{\circ}$C for 24 hours. The resulting materials were ground and isostatically pressed into the form of rods 10 cm in length and 10 mm in diameter, which were then sintered at 850 $^{\circ}$C. To limit the loss of Na during the firing steps of this process, the samples were placed in a pre-heated furnace. Single crystals were grown in a NEC two mirror infra red image furnace. The growths were carried out in 2.5 atmosphere of oxygen at a growth rate of 10 mm/hour. X-ray powder diffraction of the polycrystalline precursors and a crushed sample of the resulting crystal showed that the materials were single phase to within the resolution of the diffractometer. Crystals for the measurements were obtained by cleaving from the boule. Sodium was removed by a chemical deintercalation technique using I$_2$ in an acetonitrile solution. The x value for each sample was measured using EDX on a JSM 6100 scanning electron microscope.
\par
Magnetic susceptibility ($\chi$) measurements were carried out in a Quantum Design MPMS-5S superconducting quantum interference device (SQUID) magnetometer. Hydrostatic pressures (P) of up to 10 kbar were applied to the samples using an easyLab Mcell 10 pressure cell designed for use with this magnetometer. Crystals with masses of 20 - 30 mg were placed in a cylindrical PTFE sample holder. This container, which has an internal diameter of 1.9 mm and a length of 10 mm, was then filled with a pressure transmitting medium (Daphne oil) and placed into the sample space of the beryllium-copper piston clamp pressure cell. Pressure is applied at room temperature. Measurements were carried out at temperatures between 2 and 50 K with magnetic fields of up to 50 kOe applied parallel to the \textit{ab} planes of the crystals. The pressure cell has a single wall design giving it a total weight of only 34 g. The background signal from the cell was removed by first carrying out measurements under the same experimental conditions with an empty cell. We then used the background subtraction facility of the Quantum Design magnetometer, which corrects each measurement scan point by point, before fitting the measured response curve to determine the magnetic moment of the sample. The pressure at low temperature was determined in situ by measuring the superconducting transition temperature in an applied magnetic field of 10 Oe of a small piece of high purity (99.9999$\%$) tin wire placed alongside the sample. Heat capacity data were collected with a Quantum Design Physical Properties Measurement System (PPMS) in magnetic fields up to 90 kOe. The heat capacity (C) measurements were carried out by a two-tau relaxation method using both a standard $^4$He and a $^3$He insert. A background signal (platform and Apiezon N grease) was recorded versus temperature for each run.
\section{Results}
\begin{figure}[ht]
\centering
\includegraphics[width = 10 cm]{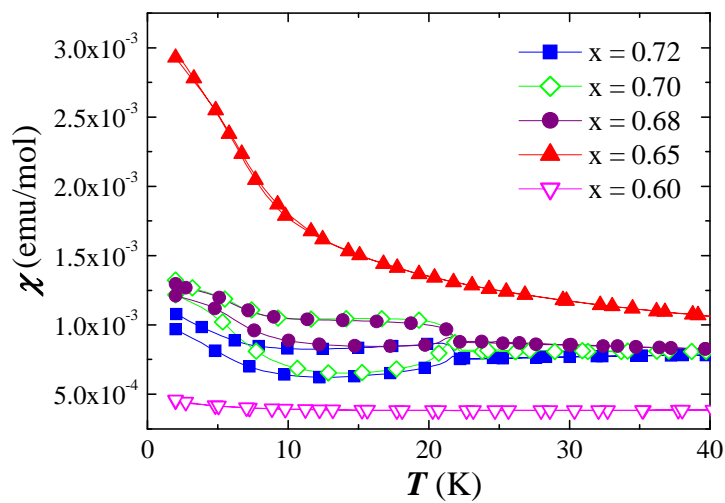}
\caption{Magnetic susceptibility versus temperature of Na$_x$CoO$_2$ with the applied field parallel to the \textit{c} axis. For 0.68$\leq$x$\leq$0.72 the transition temperature, T$_{mag}$, is invariant with x. The signature of magnetic order disappears altogether at a doping level of x=0.65 and leads to a rapid increase in $\chi$, followed by a decrease to an almost T independent susceptibility at x=0.60.}
\label{chivsx}
\end{figure} 
\par
Figure~\ref{chivsx} shows the dc magnetic susceptibility versus temperature dependence with H=1 kOe applied parallel to the \textit{c} axis for a sequence of Na$_x$CoO$_2$ materials with varying Na content. The feature at T$_{mag}$=22.0 K indicates the onset of magnetic order. As the sodium content is reduced, T$_{mag}$ remains constant. Any signature of magnetic ordering disappears abruptly at x=0.65. At temperatures below T$_{mag}$ there is considerable evolution of the magnetic response with x. The degree of hysteresis between the zero field cooled (ZFC) and field cooled cooling (FCC) curves reaches a maximum at x=0.70. There is a signature of a modification to the magnetic response at 9 K in the FCC data sets for samples with 0.65$\leq$x$\leq$0.72. The removal of Na and the eventual suppression of magnetic order is accompanied by an increase in the overall magnitude of the magnetic susceptibility. This is due to the increasing number of Co spins that are successively decoupled from the antiferromagnetic SDW; an increase in the paramagnetic signal is consistent with an increase in uncompensated moments. This is followed by a rapid reduction in $\chi$ to an almost T independent signal for x=0.60.
\par
\begin{figure}[ht]
\centering
\includegraphics[width = 10 cm]{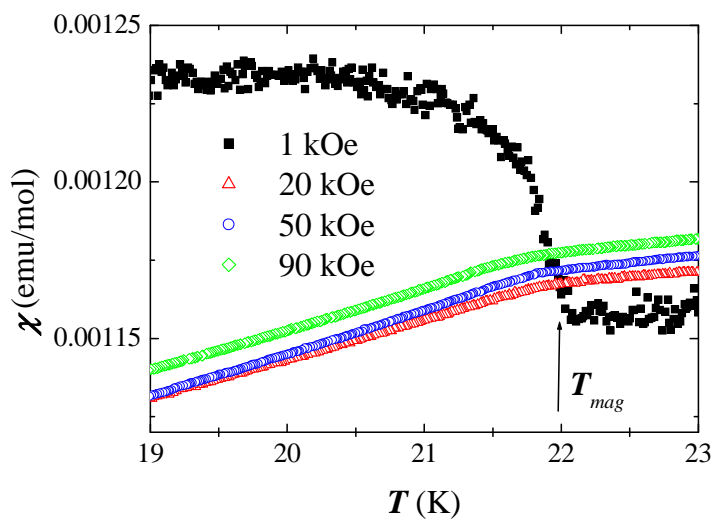}
\caption{Magnetic susceptibility versus temperature measurements for Na$_{0.70}$CoO$_2$ show that the magnetic ordering transition temperature is invariant under different applied fields (H$_\textit{ab}$$\leq$90 kOe) for this composition. The 20 kOe$\leq$H$\leq$90 kOe data sets have been offset by 5$\times$10$^{-6}$ emu/mol for clarity.}
\label{chivsh}
\end{figure}

In figure~\ref{chivsh} we show magnetization versus temperature data collected for an x=0.70 sample in fields of 1, 20, 50 and 90 kOe. T$_{mag}$ appears to remain fixed at 22.0 K for all fields.
\par
\par
\begin{figure}[ht]
\centering
\includegraphics[width = 10 cm]{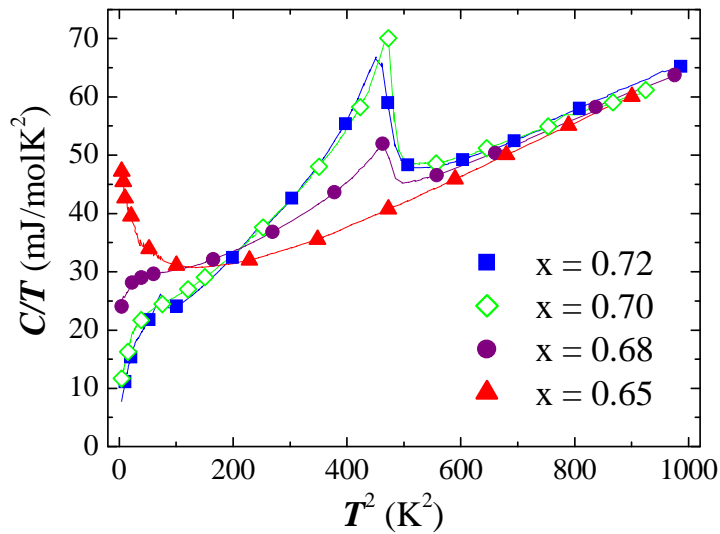}
\caption{ C/T versus T$^2$ curves for Na$_x$CoO$_2$ at different doping levels. The size of the anomaly remains constant for 0.70$\leq$x$\leq$0.72 and then decreases with decreasing sodium content while the transition temperature T$_{mag}$ remains constant. No transition is evident for x$\leq$0.65.}
\label{cvsx}
\end{figure}

C/T versus T$^2$ curves for samples with 0.65$\leq$x$\leq$0.72 are shown in figure~\ref{cvsx}. A fit to the data above T$_{mag}$ gives $\gamma$=22.5 mJ/molK$^2$. The data contain a lambda like anomaly at 22.0 K indicating the onset of magnetic ordering that is present in all materials down to x=0.68. The transition is broad extending over at least 7 K. This feature occurs at the same temperature in both heating and cooling runs with no discernable hysteresis to within the experimental accuracy of the technique used, indicating that this is a second-order phase transition. The magnitude of the anomaly in C ($\sim$0.4 J/molK) remains fixed for 0.70$\leq$x$\leq$0.72, but then gradually diminishes before disappearing at x=0.65. This jump corresponds to up to 30$\%$ of the signal at T$_{mag}$.
\par
\par
At low T, the temperature dependence of C varies enormously; $\gamma$ increases from 8 mJ/molK$^2$ at x=0.72 to approach 50 mJ/molK$^2$ at x=0.65. The reduction in $\gamma$ below T$_{mag}$ suggests that up to 50 - 60$\%$ of the Fermi surface is removed by the opening up of a gap as the system enters a magnetic ordered state. The destruction of this magnetically ordered state for lower x may leave us with a system of highly correlated electrons with an enhanced $\gamma$. Alternatively, the increase in C/T versus T at low T for x=0.65 may be the result of a Schottky anomaly. There are several features in the heat capacity data which correlate well with the anomalies seen in the magnetization data and suggest that the magnetically ordered state of this material undergoes significant modification below T$_{mag}$.
\par
\begin{figure}[ht]
\centering
\includegraphics[width = 10 cm]{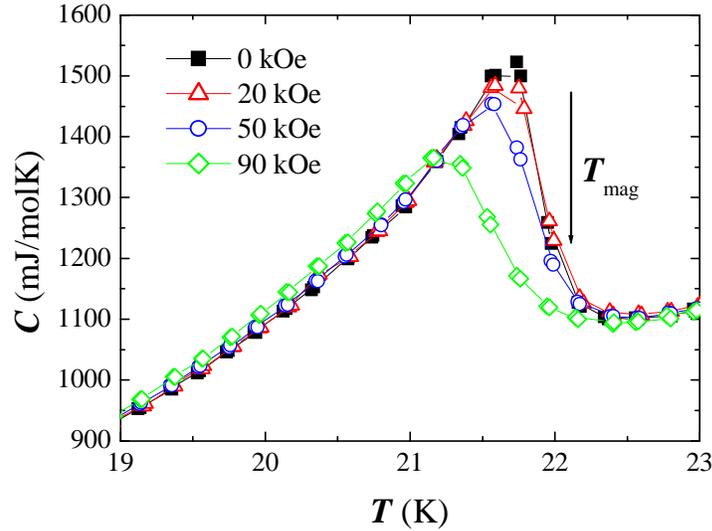}
\caption{Heat capacity of Na$_{0.70}$CoO$_2$ in various magnetic fields applied parallel to the \textit{c} axis of the crystals. The temperature of both the peak and the midpoint of the transition are reduced in field indicating a broadening with applied field, however the onset temperature remains constant.}
\label{cvsh}
\end{figure}

In figure~\ref{cvsh} we show specific heat capacity versus temperature data collected for an x=0.70 sample in fields of 0, 20, 50 and 90 kOe applied parallel to the \textit{c} axis. The application of fields of up to 90 kOe leaves the temperature at which the onset of the magnetic ordering is observed unchanged. The transition is progressively broadened and the magnitude of the anomaly around T$_{mag}$ is considerably reduced as the applied field is increased.
\par
\par
\begin{figure}[ht]
\centering
\includegraphics[width = 10 cm]{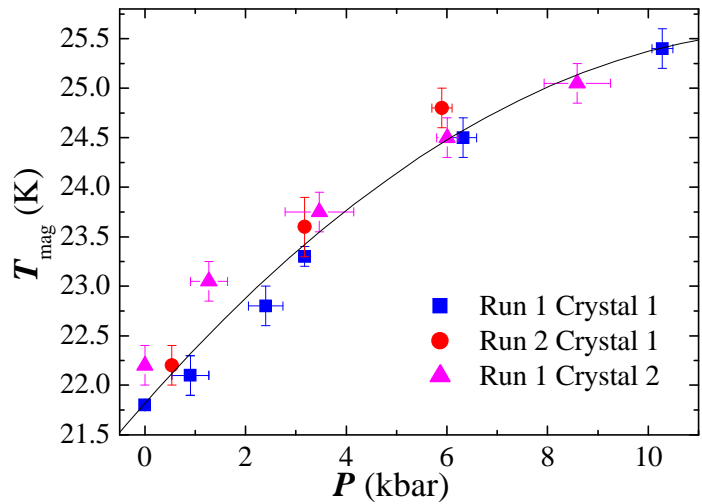}
\caption{The variation of T$_{mag}$ for two Na$_{0.7}$CoO$_2$ crystals. A clear increase in T$_{mag}$ is seen with increasing pressure. Initially, dT$_{mag}$/dP=+0.44 K/kbar with a tendency towards saturation at higher pressures. (The solid line is a guide to the eye).}
\label{pdata}
\end{figure}
\begin{figure}[ht]
\centering
\includegraphics[width = 10 cm]{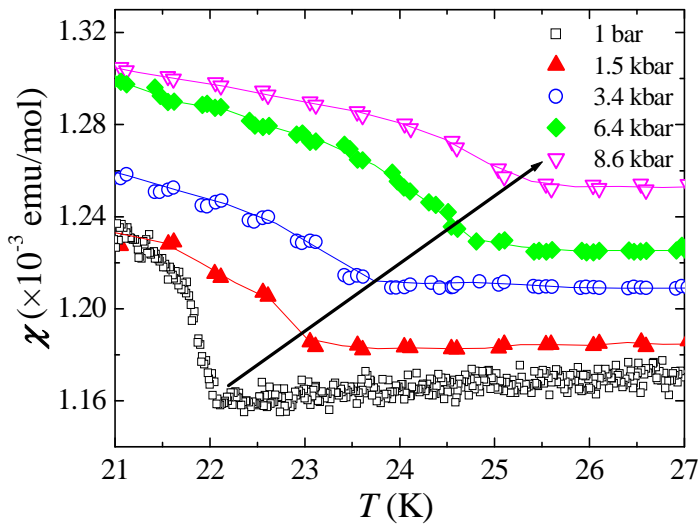}
\caption{The raw magnetic susceptibility data showing the variation in T$_{mag}$ at five different applied pressures. The data are offset vertically from each other by 2$\times$10$^{-5}$ emu/mol so that the shift of T$_{mag}$ to higher temperatures (as indicated by the arrow) is visible.}
\label{rawpdata}
\end{figure}

Figure~\ref{pdata} shows the pressure dependence on T$_{mag}$ for two single crystal samples of x=0.70. Data were collected in three separate runs in magnetic fields of 50 kOe with both increasing and decreasing pressure. The raw magnetization data for one of the crystals is shown in figure~\ref{rawpdata}. The signature of magnetic order is seen at T$_{mag}$=22.0$\pm$0.2 K at 1 bar and initially increases at a rate $\partial$T$_{mag}$/$\partial$P=+0.44$\pm$0.03 K/kbar. This increase begins to saturate at higher pressures with T$_{mag}$ reaching 25.4 K at 10 kbar.
\par
\par
\par
For similar measurements on an x=0.65 sample, (i.e. a composition for which there are no features in either the magnetization or the heat capacity data that can be associated with the onset of magnetic order) the application of pressures of up to 10 kbar does not induce any features in the M versus T data that indicate that pressure can restore the magnetic order.
\section{Discussion}
The details of the evolution of the magnetic state at temperatures below T$_{mag}$ will be discussed in a separate paper. Here we focus on the effects of magnetic field, composition, and pressure on the magnetic ordering temperature T$_{mag}$.
\par
We must first consider whether there is any equivalence between the pressures applied here and the effects of doping. There are few studies of the pressure dependence of the structure of Na$_x$CoO$_2$. Rivadulla et al.~\cite{rivadulla} have measured the change in the room temperature lattice parameters with pressure up to 45 kbar for a powder sample of x=0.57 crushed from single crystals. The structure remained unchanged, with a first order phase transition at 35 kbar separating a highly compressible low-pressure phase from a high-pressure phase which is less compressible. The \textit{c}-axis initially decreases from 10.99 \AA\, at ambient pressure to 10.88 \AA\, at 10 kbar ($\partial$\textit{c}/$\partial$P=-0.012 \AA/kbar ) whilst \textit{a} decreases from 2.8293 to 2.8193 \AA\, ($\partial$\textit{a}/$\partial$P=-0.001 \AA /kbar) over the same pressure range.
\par
Assuming a similar behaviour across the Na$_x$CoO$_2$ series we can use the data of Huang et al.~\cite{huang} to estimate that a reduction in x from x=0.71 to x=0.63 brings about a change in \textit{c} (\textit{a}) that could be induced by the application of a positive (negative) pressure of around 11 kbar (7 kbar). This suggests that the magnitude of any chemical pressure induced through doping should be comparable with our maximum externally applied hydrostatic pressure of 10 kbar, although doping will also alter the nature of the electronic configurations within the system.
\par
There have been suggestions that over the range of x considered here, Na$_x$CoO$_2$ phase segregates into magnetic and non magnetic regions~\cite{carretta}. This may explain the gradual reduction in the magnitude of the jump at T$_{mag}$ seen in the C(T) data and if there is a non magnetic phase, why the application of pressure is unable to restore the magnetic order. Whether this is the case or not, it is clear that the application of moderate pressure modifies the exchange pathways within the magnetically ordered phase sufficiently to strongly influence the magnetic order. We must therefore consider how this may occur and what consequences this has for our understanding of the nature of the magnetic order in this material.
\par
How does the observed behaviour compare with that of well characterised SDW systems? In both the 1D Bechgaard salts~\cite{biskup} and the 2D (ET)$_2$X organic conductors~\cite{brooks}, an almost perfect nesting of the Fermi surface stabilises a SDW state. The application of pressure increases the deviation from perfect nesting and suppresses the SDW. In other words, pressure increases the dimensionality of the system, which destroys the Fermi surface topology necessary to support the SDW ground state. For a perfect nesting case~\cite{hanasaki}, the application of a magnetic field also decreases the SDW transition temperature; in contrast, for imperfect nesting, the SDW ordering temperature increases in an applied field. 
\par
The behaviour reported here is different, with the application of pressure leading to an increase in T$_{mag}$, while T$_{mag}$ is fixed in an applied field. The notion that a SDW state in Na$_x$CoO$_2$ could be stabilised by a decrease in the dimensionality of the system is also at odds with the experimental observation that while the \textit{c}/\textit{a} ratio increases from $\sim$3.84 for x=0.71 to $\sim$3.9 at x=0.64~\cite{huang}, since T$_{mag}$ initially remains constant before magnetic order disappears as x decreases over this x range.
\par
In other systems more closely related to Na$_x$CoO$_2$ however, it has been shown that the response of the magnetically ordered state to the application of pressure reflects the details of the crystal structure and cannot be understood by simply assuming that the application of hydrostatic pressure leads to a uniform compression along the different crystallographic directions. In the 3 and 4-layer cobaltites [Ca$_2$CoO$_3$]$_{0.62}$[CoO$_2$] and [Ca$_2$Co$_{4/3}$Cu$_{2/3}$O$_4$]$_{0.62}$[CoO$_2$], muon spin rotation and relaxation experiments suggest that the transition to an incommensurate spin density wave state is independent of pressure up to 13 kbar~\cite{sigiyama2}. In these materials, the CoO$_2$ planes are separated by rocksalt (RS) type blocks. The authors of this study speculate that the application of pressure induces large changes in the Co-O distances within the RS-type blocks and that if the incommensurate spin density wave (IC-SDW) exists in the CoO$_2$ plane little change in the nature of this magnetic feature is expected.
\par
Does such a scenario apply here? Once again the data available are limited. We know that at room temperature the thickness of the CoO$_2$ layer and the Co-O distance remain almost constant for 0.7$\leq$x$\leq$0.76, while the thickness of the NaO$_2$ layer increases~\cite{huang}. If it is principally the thickness of the CoO$_2$ layer that controls the magnetism, this may explain why T$_{mag}$ remains constant within this range of x and why the application of pressure leads to an increase in T$_{mag}$. However, given that the CoO$_2$ layer thickness continues to decrease with x for x$\leq$0.7 while the signature of magnetic order disappears, we must assume that the thickness of the NaO$_2$ layer or other factors also play a role. For instance, in the local moment representation, each Na donates an electron to the CoO$_2$ network, altering the formal valence on the cobalt to be 3+(1-x). Recent calculations~\cite{johannes} of the exchange pathways within this material suggest that superexchange via direct O-O hopping and/or through intermediate Na atoms (i.e. Co-O-Na-O-Co) are important and that the exchange integrals are rather more two dimensional than suggested by the neutron scattering data~\cite{bayrakci,helme}. This interplanar Co-Co AFM coupling, whether mediated through the Na orbitals or not, will undoubtedly rely on the Na occupation level. Since superexchange depends more strongly on distance in comparison to the in-plane FM double exchange interactions, a decrease in \textit{c} caused by applying external pressure will result in a marked change in inter-planar hopping distances without altering number of superexchange pathways. In order to ascertain whether changes in the CoO$_2$ plane play any significant role, detailed structural studies as a function of pressure are necessary.
\section{Summary}
For Na$_x$CoO$_2$, the chemical pressure induced through doping in the composition range 0.6$\leq$x$\leq$0.75 allows a meaningful comparison with measurements made under moderate hydrostatic pressures (10 kbar). Whether or not the system is phase segregated, we have shown that pressure can significantly modify the magnetic ordering temperature, T$_{mag}$, of Na$_x$CoO$_2$ for x$\sim$0.7. The observed pressure dependence of T$_{mag}$ does not comply with the behaviour expected for many well studied SDW systems. This suggests that either the magnetic order does not take the form of a spin density wave or that the application of pressure produces Fermi surface effects that are not typical of model SDW systems. Our results highlight the need for detailed measurements of the temperature and pressure dependence of the structure of Na$_x$CoO$_2$ at temperatures directly above and below T$_{mag}$.
\ack
We acknowledge the financial support of the EPSRC (UK) for this project. We would also like to thank Dr Christophe Thessieu and easyLab for providing the Mcell10 pressure cell and for help with the measurements made under pressure.

\end{document}